\documentclass[a4paper,11pt]{article}
\usepackage{pos}

\title{Topological susceptibility and QCD phase transition with 2+1 flavor M\"obius domain wall fermion at finite temperature}
\ShortTitle{Topological susceptibility and QCD pahse transition with 2+1 flavor M\"obius...}

\author*[a]{Issaku Kanamori}
\author[a]{Yasumichi Aoki}
\author[b]{Hidenori Fukaya}
\author[c]{Jishnu Goswami}
\author[d,e]{Shoji Hashimoto}
\author[c]{Yu Zhang}
\onbehalf{ (JLQCD collaboration)}

\affiliation[a]{RIKEN Center for Computational Science (R-CCS), %
Kobe 650-0047, Japan}

\affiliation[b]{Department of Physics, Osaka University, %
Toyonaka, Osaka 560-0043, Japan}

\affiliation[c]{Fakult\"at f\"ur Physik, Universit\"at Bielefeld, Bielefeld, 33615, Germany}

\affiliation[d]{KEK Theory Center, High Energy Accelerator Research Organization (KEK), %\\
Tsukuba 305-0801, Japan}

\affiliation[e]{School of High Energy Accelerator Science, Graduate University for Advanced Studied (SOKENDAI),\\
Tsukuba 305-0801, Japan}

\emailAdd{kanamori-i@riken.jp}
\emailAdd{yasumichi.aoki@riken.jp}
\emailAdd{hfukaya@het.phys.sci.osaka-u.ac.jp}
\emailAdd{jishnu@physik.uni-bielefeld.de}
\emailAdd{shoji.hashimoto@kek.jp}
\emailAdd{yzhang@physik.uni-bielefeld.de}

\abstract{%
The topological susceptibility is one of the quantities that has a large discretization error, and the error can be sensitive to the choice of fermion action. We report on our results from physical point simulations with 2+1 flavor M\"obius domain wall fermion at finite temperature.
We also present the chiral condensate and disconnected susceptibility.
The temporal lattice size is $N_t=12$ and $16$, and the temperature range is around 140 MeV to 250 MeV
for the chiral condensate and susceptibility.  A coarse lattice with $N_t=10$ covers up to 500 MeV to measure the topological susceptibility. 
}

\FullConference{The 42nd International Symposium on Lattice Field Theory (LATTICE2025)\\
2-8 November 2025\\
Tata Institute of Fundamental Research, Mumbai, India\\}

\begin{document}
\maketitle

\section{Introduction}

The chiral symmetry is the most important symmetry in QCD.
It is therefore desirable to keep the chiral symmetry as much as possible
in the lattice simulation.  M\"obius domain wall fermion (MDWF) is an optimal
choice of the lattice fermion that keeps a balance between the chiral symmetry
and the computational cost.
In this talk, we use 2+1 flavor MDWF fermion at the physical point at finite temperature
and present our preliminary results of chiral condensate/susceptibility and topological susceptibility.

Topological susceptibility is one of the fundamental quantities that characterizes
the vacuum of QCD.  The behavior at high temperature also plays an important role in the axion dark matter scenario.
However, it is known that it may have a large discretization error
depending on the discretizations of fermions.
We need to quantify the discretization effects as well as possible systematic errors,
which is especially challenging at high temperature.
Because of the topology freezing at high temperature, sampling over topological sectors becomes difficult.  We also need to use a very fine lattice to resolve the
temporal direction.
The results reported so far \cite{Bonati:2015vqz,Petreczky:2016vrs,Borsanyi:2016ksw, Chen:2022fid,Kotov:2025ilm, Athenodorou:2022aay,Gavai:2024mcj}
in fact show a large discrepancy among
the collaborations.  Even among the results with the domain wall fermion, TWQCD reported a significantly larger value at high temperature than
other collaborations with their optimal domain wall simulation \cite{Chen:2022fid}.
As reported by Gavai et al.~\cite{Gavai:2024mcj}, M\"obuis domain wall fermion 
has small discretization errors, of which the detailed simulation setting is different from ours.  They showed that $N_t=8$ lattice result is already
very close to the continuum limit with HISQ fermion.
A new result with overlap fermion \cite{Fodor:2025mqi, Kotov:lattice2025} with $N_t=8$ has appeared recently.

We also present the chiral condensate and disconnected susceptibility\footnote{%
We originally planned to deliver the results of these quantities by Y.Z. as an independent talk,
but had not realized 
because she could not attend the conference on site.
}. 
They are, of course, important quantities to investigate the restoration of chiral symmetry
at high temperature.
Keeping a good chiral symmetry with MDWF helps to control systematic errors caused by 
the lattice artifact that breaks the chiral symmetry.
With finite fifth-dimensional extent, the quark mass receives an additive contribution known as the residual mass, which contributes to the additive divergence that must be removed during renormalization.
The result we present in this contribution implies that we have good control of
the residual mass correction. Furthermore, the crossover transition temperature at the physical point read off from the peak location of the chiral susceptibility is consistent with results obtained from the HISQ and stout actions in the continuum limit~\cite{HotQCD:2018pds,Borsanyi:2020fev}.

After explaining the simulation setting in the next section,
we first present the result on chiral condensate and disconnected susceptibility in Sec.~\ref{sec:pbp}. 
Then, we show the behavior of topological charge and susceptibility in Sec.~\ref{sec:top_susc}.  Section~\ref{sec:summary} is for the summary and outlook.

\section{Simulation Setup}
\label{sec:setup}

We use M\"obius domain-wall fermion with the scale factor 2
and gauge action with tree-level Symanzik-improvement.
For the gauge field inside the fermion action, we apply 3 steps of stout smearing with the smearing parameter $\rho=0.1$.  
The details of the action used in the JLQCD collaboration are found in the supplement material of \cite{Colquhoun:2022atw}.

The quark masses are set to the physical point along the line of constant physics, where the renormalized light quark mass $m^R_l$ is set to $1/27.4$ of the strange quark mass $m^R_s$ at the renormalization scale $\mu=2$ GeV.  The dimensionless input masses, $m^{\text{latt}}_l$ and $m^{\text{latt}}_s$, are carefully chosen by subtracting the effect of the residual mass correction \cite{Aoki:2021kbh, Goswami:2025euh}. The fifth-dimensional extent $L_s$ is fixed to $12$ for all ensembles.
We have two different lattice spacings with two different lattice volumes for
$145 \text{\ MeV} \leq  T \leq 250 \text{\ MeV}$: $36^3\times 12$ (p2 series) and $48^3\times 12$ (p3) for the temporal lattice size $N_t=12$, $48^3\times 16$ (q2) and $64^3\times 16$ (q3) for $N_t=16$.  Some of these parameters
cover $130$ MeV in the low temperature side, and $300$ MeV in the high temperature side.
For temperature $T \geq 250$ MeV, we also have $N_t=10$ ensembles with lattice size $40^3 \times 10$ up to $T=500$ MeV.

Both configuration generations and measurements
were performed on Supercomputer Fugaku.
We used Grid \cite{Boyle:2015tjk, Meyer:2021uoj} for configuration generation.  The chiral condensate and susceptibility were measured with Grid/Hadrons \cite{antonin_portelli_2020_4293902}, and the measurement of topological charge
was performed using Bridge++ \cite{Ueda:2014rya}.

\section{Chiral Condensate and Susceptibility}
\label{sec:pbp}

The chiral condensate serves as the order parameter for chiral symmetry breaking in QCD. At finite quark mass, it suffers from both additive and multiplicative divergences. To ensure it is well defined in the continuum limit, appropriate renormalization is required. The additive divergence behaves as $C^D (m + x m_{\rm{res}})/a^2$, where the contribution $x m_{\rm{res}}/a^2$ arises from the finite size of the domain wall fermion, which induces a similar effect as the mass term does. Here, $x=\mathcal{O}(1)$ is a coefficient whose determination, along with that of $C^D$ is from our 3-flavor MDWF study and detailed in~\cite{Zhang:2025vns}. The multiplicative divergence is removed by dividing the mass renormalization constant $Z_m$ in the $\overline{\mathrm{MS}}$ scheme, obtained from a next-to-next-to-leading order running~\cite{Chetyrkin:1999pq}.

\begin{figure}
\center
\includegraphics[width=0.48\linewidth]{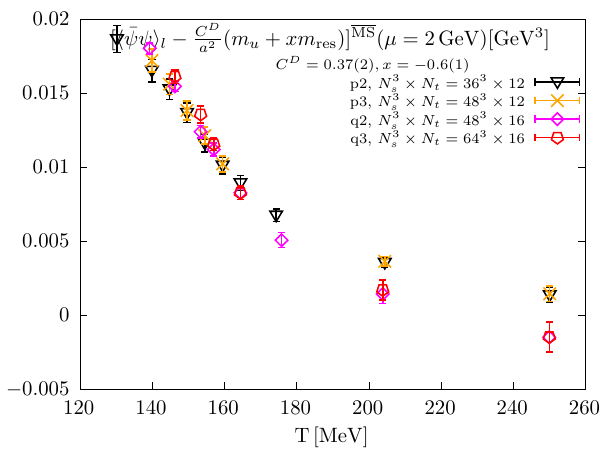}%
\includegraphics[width=0.48\linewidth]{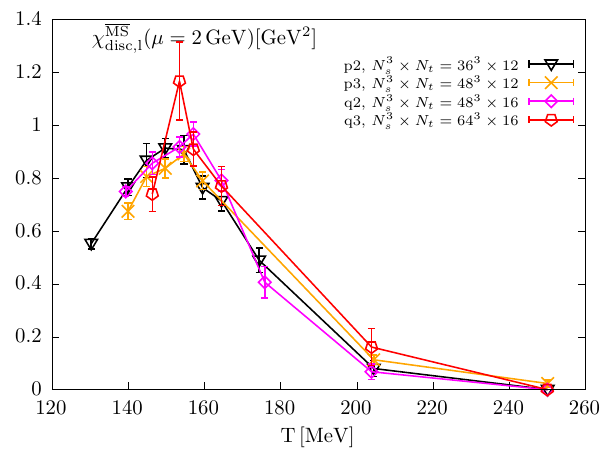}

\caption{Chiral condensate of light quarks after multiplicative and additive renormalization (left) and chiral susceptibility after multiplicative renormalization.}
\label{fig:pbp_and_susc}

\end{figure}

The left panel of Fig.~\ref{fig:pbp_and_susc} presents the renormalized light quark chiral condensate at the physical quark mass as a function of temperature for two different volumes (with aspect ratios 3 and 4) on each of the $N_t=12$ and $N_t=16$ lattices. As expected, the chiral condensate decreases as temperature increases and eventually approaches zero at high temperatures where the chiral symmetry is restored. It is very challenging to locate the inflection point directly from the chiral condensate. For this purpose, we turn to the disconnected chiral susceptibility, $\chi_{\rm{disc}}$.

The disconnected chiral susceptibility quantifies the fluctuations of the chiral condensate. By definition, its construction as a variance ensures that the additive divergences cancel, leaving only a multiplicative divergence, which is removed by dividing by $(Z_m^{\overline{\mathrm{MS}}})^2$. The right panel of  Fig.~\ref{fig:pbp_and_susc} shows the renormalized disconnected chiral susceptibility as a function of temperature for $N_t=12$ and $N_t=16$, each with two different volumes. A pronounced peak, defining the pseudocritical temperature, is observed in the range of approximately 153-157 MeV. The chiral susceptibility does not exhibit significant volume or discretization effects, except in the q3 series. It should be noted that the statistics for the first three data points of the q3 series are around 300-400 configurations, substantially fewer than the approximately 2000 configurations for the other series. This explains the larger error bars for q3, particularly at the peak. Our estimate is more or less consistent with results obtained in the continuum limit using the HISQ and stout actions, which are 156.5(1.5) MeV~\cite{HotQCD:2018pds} and 158.0(6) MeV~\cite{Borsanyi:2020fev}, respectively.
It is also consistent with other MDWF studies with $N_t=8$ ensembles, which report 155(9) MeV~\cite{Bhattacharya:2014ara} and $158.7^{+2.6}_{-2.3}$ MeV~\cite{Gavai:2024mcj}. 

\section{Topological Susceptibility}
\label{sec:top_susc}

We use a clover discretization of the topological charge after applying Wilson flow on the gauge field.
The flow time is fixed to 5.0 in the lattice unit, which is intended not to introduce a new scale by hand in the physical unit.
We have 9 Monte Carlo streams for each parameter set, and measure the topological charge every 10 trajectories after thermalization.  Most of the data points have around 2000 measurements or more, except for the $64^3\times 16$ lattice case.  For this largest lattice volume, available configurations are still limited, and we use only about 300-400 configurations for $T \lesssim 160$ MeV.

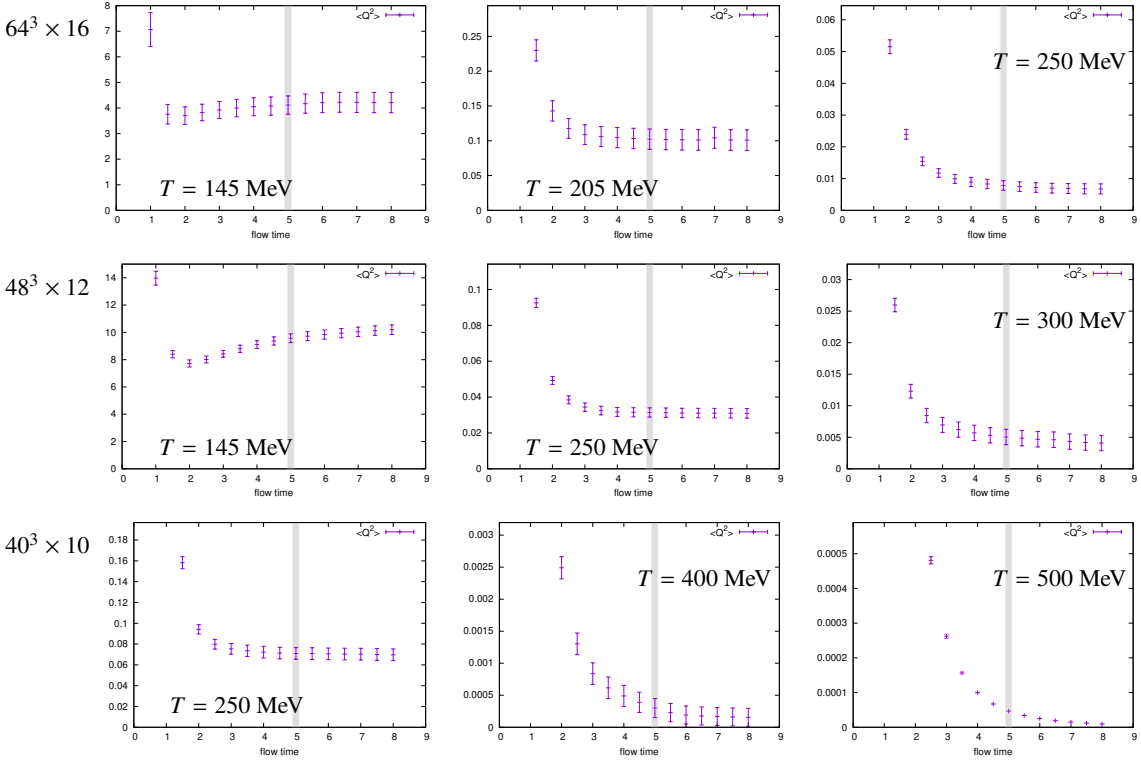
\begin{figure}
 \noindent\raisebox{0.18\linewidth}{\footnotesize $64^3 \times 16$}%
\includegraphics[height=0.215\linewidth]{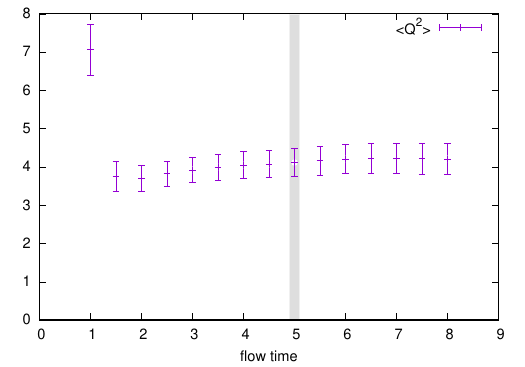}%
\includegraphics[height=0.215\linewidth]{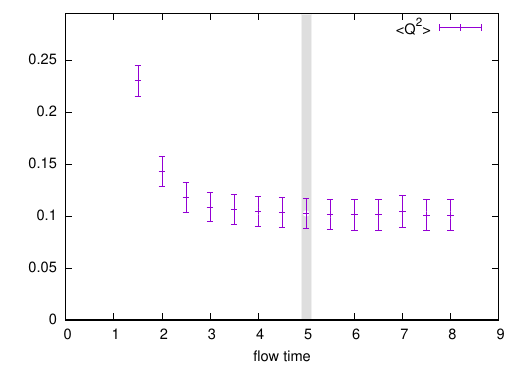}%
\includegraphics[height=0.215\linewidth]{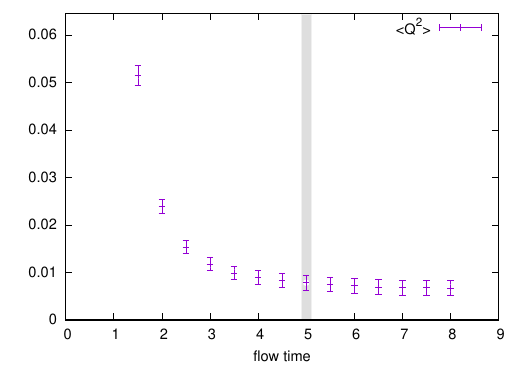}\\
\begin{picture}(0,0)(0,0)
 \put(58,32){\makebox{\footnotesize $T=145$ MeV}}
 \put(195,32){\makebox{\footnotesize $T=205$ MeV}}
 \put(370,80){\makebox{\footnotesize $T=250$ MeV}}
\end{picture}
\vspace*{-1em}

\noindent\raisebox{0.18\linewidth}{\footnotesize $48^3 \times 12$}%
\includegraphics[height=0.215\linewidth]{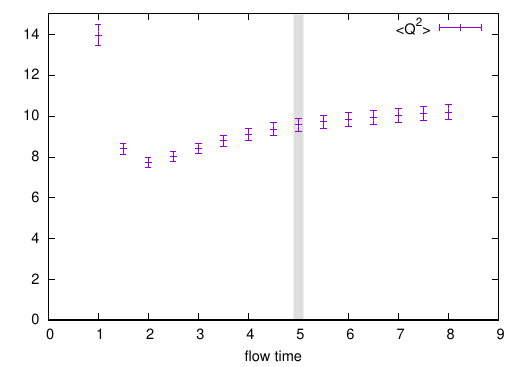}%
\includegraphics[height=0.215\linewidth]{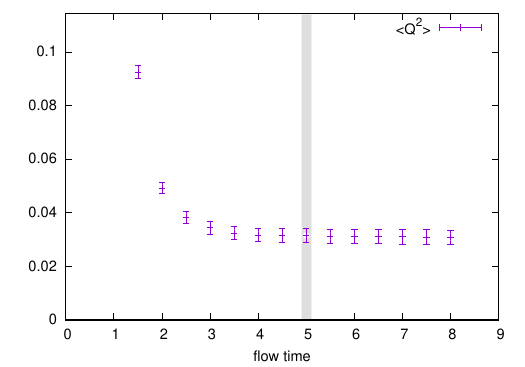}%
\includegraphics[height=0.215\linewidth]{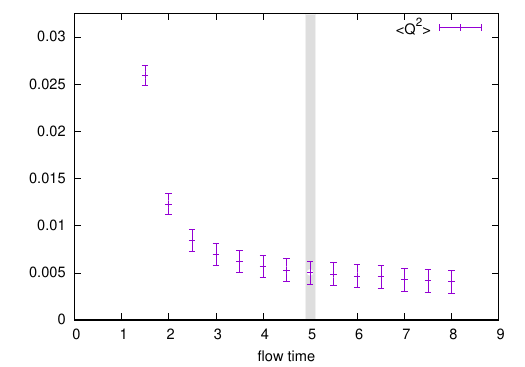}\\
\begin{picture}(0,0)(0,0)
 \put(58,32){\makebox{\footnotesize $T=145$ MeV}}
 \put(195,32){\makebox{\footnotesize $T=250$ MeV}}
 \put(370,80){\makebox{\footnotesize $T=300$ MeV}}
\end{picture}
\vspace*{-1em}

\noindent\raisebox{0.18\linewidth}{\footnotesize $40^3 \times 10$}%
\includegraphics[height=0.215\linewidth]{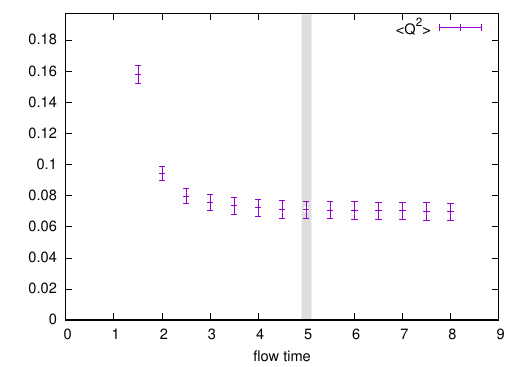}%
\includegraphics[height=0.215\linewidth]{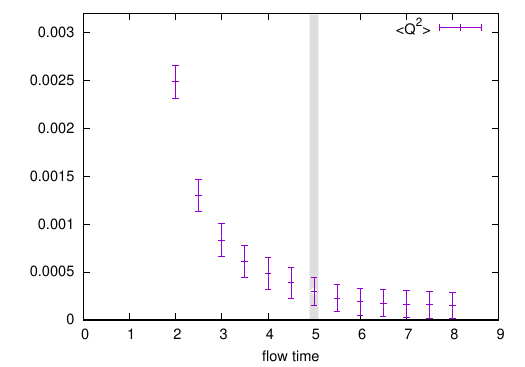}%
\includegraphics[height=0.215\linewidth]{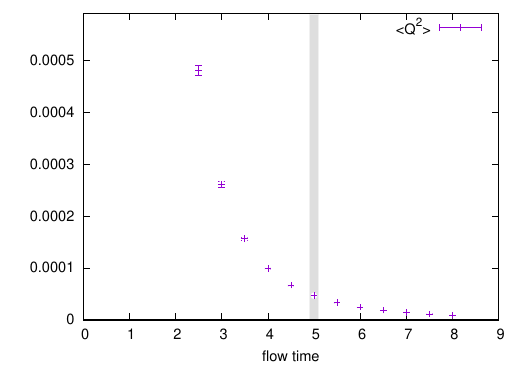}\\
\begin{picture}(0,0)(0,0)
 \put(62,32){\makebox{\footnotesize $T=250$ MeV}}
 \put(237,80){\makebox{\footnotesize $T=400$ MeV}}
 \put(370,80){\makebox{\footnotesize $T=500$ MeV}}
\end{picture}

\caption{Flow time dependence of the topological charge squared.}
\label{fig:tc2_flow_dep}

\end{figure}

\begin{figure}
 \noindent\raisebox{0.16\linewidth}{\footnotesize $64^3 \times 16$}%
\includegraphics[height=0.215\linewidth]{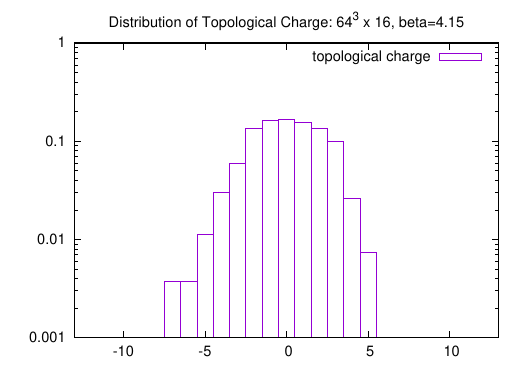}%
\includegraphics[height=0.215\linewidth]{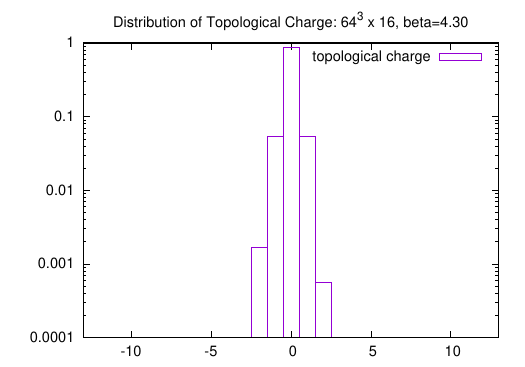}%
\includegraphics[height=0.215\linewidth]{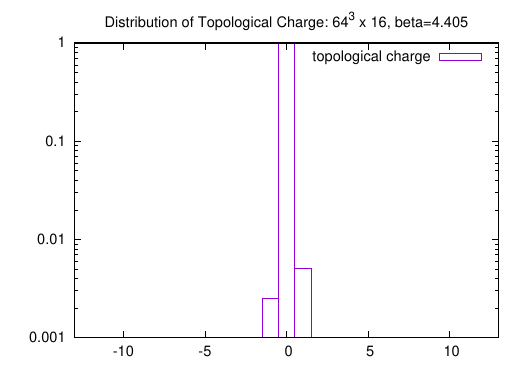}\\
\begin{picture}(0,0)(0,0)
 \put( 55,82){\makebox{\footnotesize $T=145$ MeV}}
 \put(190,82){\makebox{\footnotesize $T=205$ MeV}}
 \put(321,82){\makebox{\footnotesize $T=250$ MeV}}
 \put( 55,71){\makebox{\footnotesize \color{red}223/268}}
 \put(190,71){\makebox{\footnotesize \color{red}199/1789}}
 \put(321,71){\makebox{\footnotesize \color{red}15/1969}}
\end{picture}
\vspace*{-1em}

\noindent\raisebox{0.16\linewidth}{\footnotesize $48^3 \times 12$}%
\includegraphics[height=0.215\linewidth]{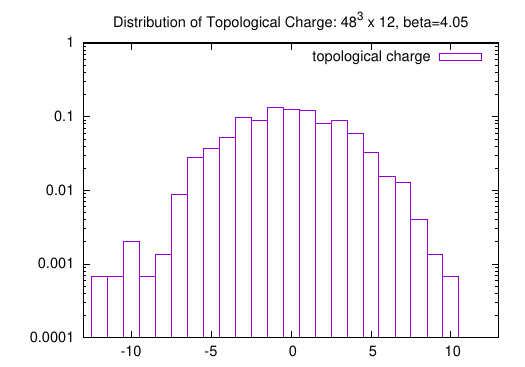}%
\includegraphics[height=0.215\linewidth]{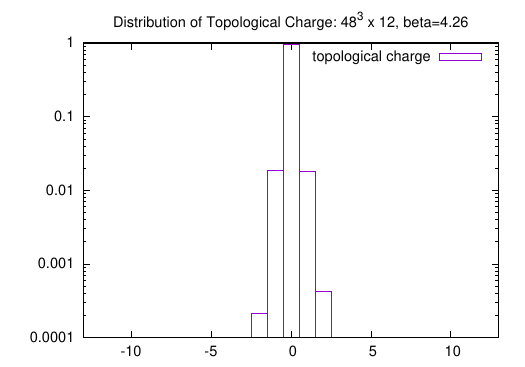}%
\includegraphics[height=0.215\linewidth]{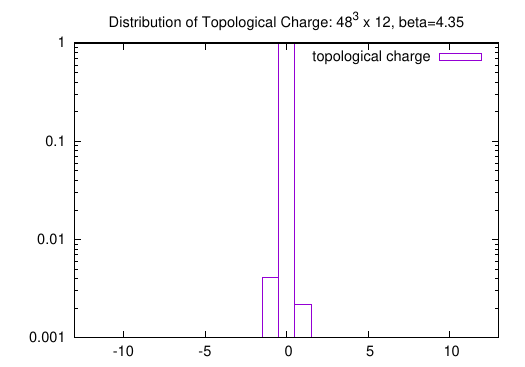}\\
\begin{picture}(0,0)(0,0)
 \put( 55,82){\makebox{\footnotesize $T=145$ MeV}}
 \put(190,82){\makebox{\footnotesize $T=250$ MeV}}
 \put(321,82){\makebox{\footnotesize $T=300$ MeV}}
 \put( 55,71){\makebox{\footnotesize \color{red}1283/1467}}
 \put(190,71){\makebox{\footnotesize \color{red}176/4689}}
 \put(321,71){\makebox{\footnotesize \color{red}29/4613}}
\end{picture}
\vspace*{-1em}

\noindent\raisebox{0.16\linewidth}{\footnotesize $40^3 \times 10$}%
\includegraphics[height=0.215\linewidth]{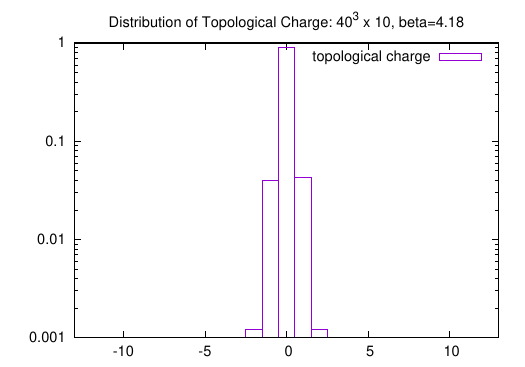}%
\includegraphics[height=0.215\linewidth]{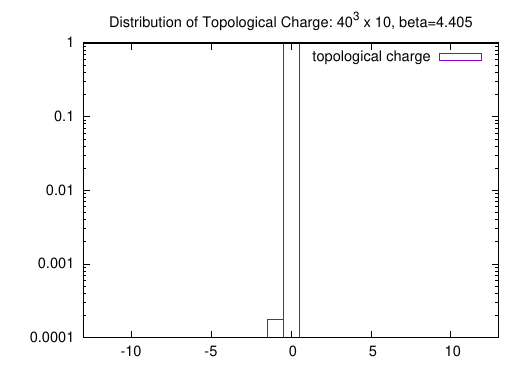}%
\includegraphics[height=0.215\linewidth]{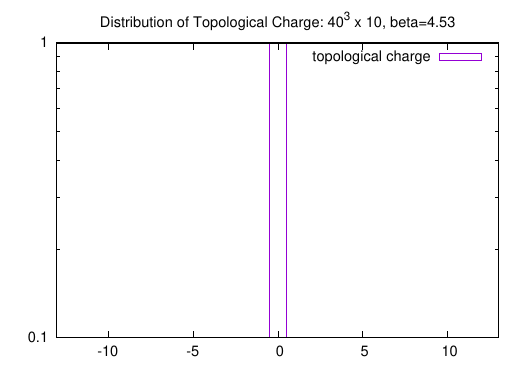}\\
\begin{picture}(0,0)(0,0)
 \put( 55,82){\makebox{\footnotesize $T=250$ MeV}}
 \put(190,82){\makebox{\footnotesize $T=400$ MeV}}
 \put(321,82){\makebox{\footnotesize $T=500$ MeV}}
 \put( 55,71){\makebox{\footnotesize \color{red} 211/2469}}
 \put(190,71){\makebox{\footnotesize \color{red} 1/5649}}
 \put(321,71){\makebox{\footnotesize \color{red} 0/5109}}
\end{picture}

\caption{Distribution of the topological charge.  The numbers denoted in red in each panel refer to the number of configurations with non-zero charges after rounding to an integer, and the measured configurations.}
\label{fig:tc2_dist}

\end{figure}
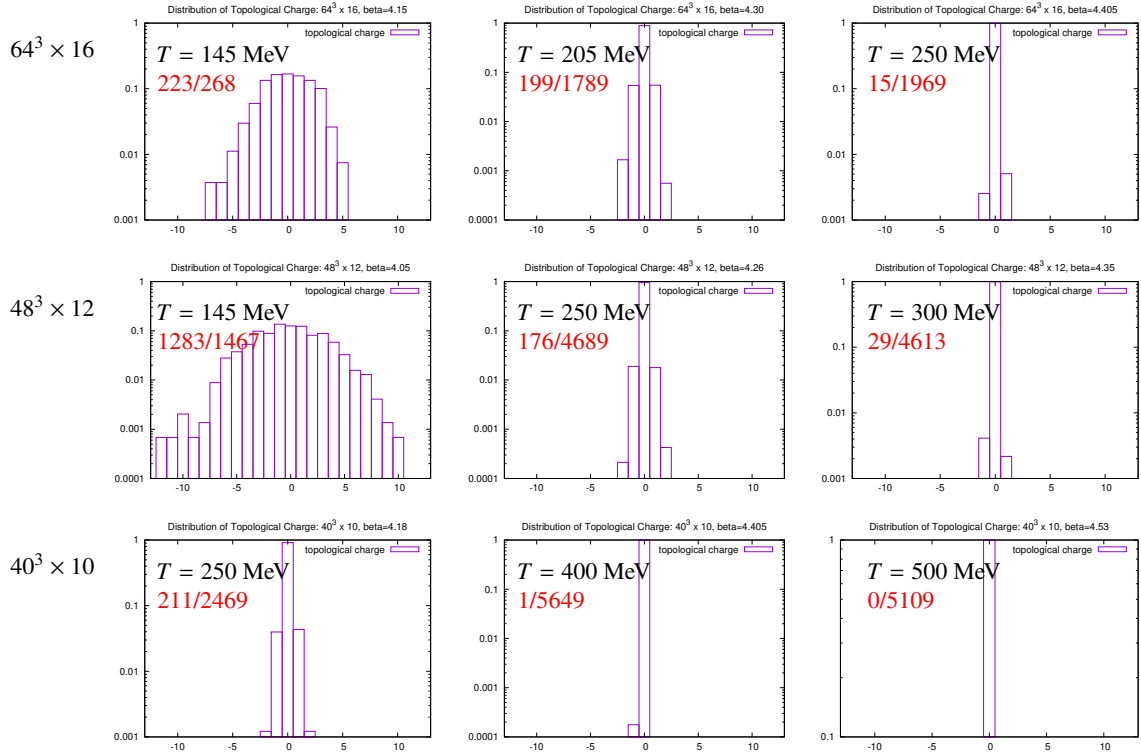

In Figure~\ref{fig:tc2_flow_dep}, we plot typical examples of the flow time dependence
of the topological charge squared. The vertical gray line indicates the flow time we adopt.
Except for very high temperatures $T\geq 400$ cases, the value stays in the plateau region. 
Figure~\ref{fig:tc2_dist} shows the distributions of the topological charge
$Q$ for the same examples.  The distributions at $T\geq 400$ are concentrated
to $Q=0$, where the value of $Q$ is rounded into an integer.
Especially, at $T=500$ MeV, we observed only $Q=0$ configurations. 
In evaluating the topological susceptibility $\chi_{\text{top}} = \langle Q^2 \rangle /V$, where $V$ is the four-dimensional volume, we do not round $Q$ into an integer.

\begin{figure}
\center
 \includegraphics[width=0.7\linewidth]{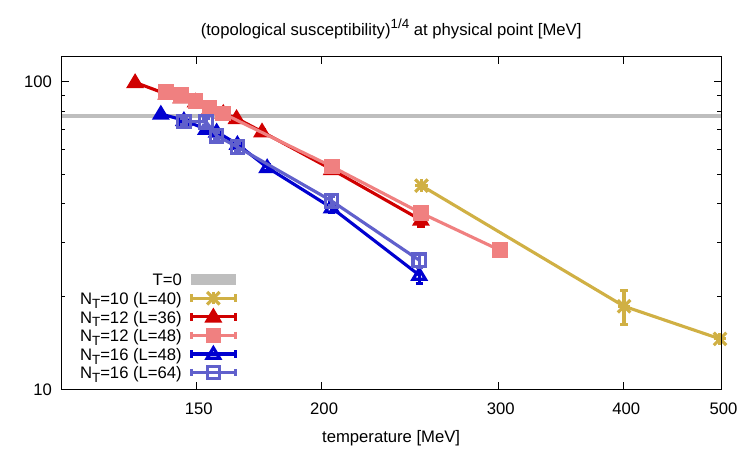}

\caption{Temperature dependence of the 4th root of topological susceptibility. From coarse lattice to fine lattice: yellow ($N_t=10$), red ($N_t=12$), and blue ($N_t=16$). The $T=0$ value is taken from \cite{Aoki:2017paw}.
}
\label{fig:sucs_top_14}

\end{figure}

The temperature dependence of the obtained topological susceptibility is 
plotted in Fig.~\ref{fig:sucs_top_14}.
The finer lattice results systematically give smaller values than the coarser lattice.
In the low temperature,
the results with $N_t=12$ overshoot the $T=0$ result taken from \cite{Aoki:2017paw},
plotted as a gray band in the figure.
This indicates that we have a sizable discretization effect at $N_t=12$.
In the low temperature phase, $T\leq 150$ MeV, we do not observe the volume dependence
within the error.
It is worth noting that our result at $N_t=16$ is closer to the continuum limit
in \cite{Borsanyi:2016ksw} with 2+1+1 flavor HISQ fermion than their $N_t=16$ result.
Their result at $T=140$ MeV is $\chi_{\text{top}}^{1/4} \approx 72$ MeV in the continuum limit and 121 MeV with $N_t=16$ calculated without rounding to an integer and eigenvalue reweighting, while our result with $N_t=16$ is 78 MeV and 74 MeV at $T=138$, $145$ MeV, respectively.

\begin{figure}
\center
\includegraphics[width=0.7\linewidth]{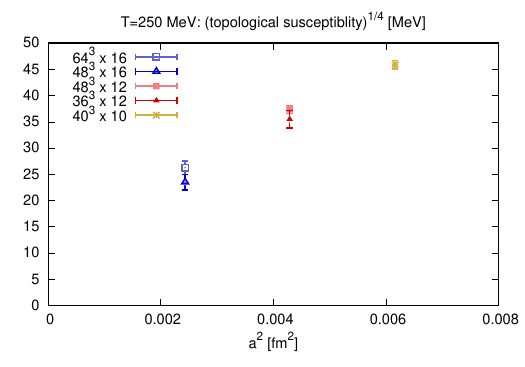}

\caption{The lattice spacing dependence of the topological susceptibility at $T=250$ MeV.
}
\label{fig:sucs_top_14_vs_a2}

\end{figure}

At temperature $T=250$ MeV, we have data from all three lattice spacings, $N_t=10$, $12$, and $16$.
Figure~\ref{fig:sucs_top_14_vs_a2} suggests that a continuum extrapolation with linear in $a^2$ should be reasonable.    
The plot also implies that even at $N_t=12$, there is a sizable finite-$a$ effect.
We reserve the systematic analysis of the continuum limit for future work.
Another observation from this plot is that the finite volume effect can be relevant.
The results with the aspect ratio 3 (triangle symbols)
are smaller than those with the aspect ratio 4 (square symbols),
although the size of the errors does not exclude the possibility of statistical
fluctuation.

\section{Summary and Outlook}
\label{sec:summary}

We presented chiral condensate and disconnected susceptibility at finite temperature with physical point M\"obius domain wall fermion with $N_t=12$ and $16$.  The preliminary result of the pseudo-critical temperature is approximately 153-157 MeV.
In the measurement of the topological susceptibility $\chi_{\text{top}}$, 
we also combined $N_t=10$ configurations
at the high temperature end.
Our result implies that the discretization error of $\chi_{\text{top}}$ with M\"obius domain wall fermion seems smaller than other result with HISQ fermion,
and the $N_t=16$ result might be close to the continuum limit. 

The simulation is still ongoing, and all the results presented here are preliminary.
We will have more statistics, especially with the $64^3 \times 16$ lattice.
The measurement of topological susceptibility at $T\geq 400$ MeV suffers from topology freezing, which requires analysis rather than simply measuring the global topological charge.
Finally, we need careful comparisons with the results from different lattice fermions.
%\subsection*{Acknowledgments}

\acknowledgments
The project is supported by the MEXT as ``Simulation for basic science: approaching the quantum era'' (JPMXP1020230411) and Joint Institute for Computational Fundamental Science (JICFuS).
We used Supercomuputer Fugaku at RIKEN Center for Computational Science (R-CCS)
for numerical simulation (HPCI project hp230207, hp240295, hp250224 
 and Usability Research ra000001).
Y. Zhang and J. Goswami acknowledge support from the Deutsche Forschungsgemeinschaft (DFG) through the CRC-TR 211 ``Strong-interaction matter under extreme conditions'' (Project No. 315477589 - TRR 211).

\bibliographystyle{JHEP}
\bibliography{proc_lattice2025.bib}

\end{document}